\documentclass[epj,pre]{svjour}

\usepackage{graphicx}

\begin{document}

\title{ Entropic uncertainty measure for fluctuations
in two-level electron-phonon models}

\titlerunning{Entropic uncertainty measure}

\authorrunning{E.Majern\'{\i}kov\'a et al.}

\author{ Eva Majern\'{\i}kov\'a\inst{1},\inst{2}  
\and 
V.Majern\'{\i}k\inst{3} 
\and S. Shpyrko\inst{1} }

\institute{Department of Theoretical Physics, Palack\'y University, \\
T\v r. 17. listopadu 50, CZ-77207 Olomouc, Czech Republic,
\email{majere@prfnw.upol.cz} \\
\and
Institute of Physics, Slovak Academy of Sciences, \\
D\'ubravsk\'a cesta, SK-84 228 Bratislava, Slovak Republic\\
\and
 Institute of Mathematics, Slovak Academy of Sciences, \\
\v Stef\'anikova 49, SK-814 73 Bratislava, Slovak Republic\\
}


\abstract{
Two-level electron-phonon systems with reflection symmetry 
linearly coupled to one or two phonon modes
(exciton and E$\otimes(b_1+b_2)$ Jahn-Teller model)
 exhibit strong enhancement of quantum fluctuations
 of the phonon coordinates and momenta
due to the complex interplay of quantum fluctuations and nonlinearities
 inherent to the models.
  We show that for the complex correlated quantum fluctuations of the
 anisotropic two-level systems the Shannon entropies  of phonon
 coordinate and momentum and their sum yield their proper global description.
On the other hand, the variance measures of the
Heisenberg uncertainties suffer from several shortcomings to provide proper
description of the fluctuations.
Wave functions, related entropies and variances
 were determined by direct numerical simulations.
Illustrative variational calculations were performed to demonstrate
the effect on an analytically tractable exciton model.
}   

\maketitle

\section{Introduction}

In spite of long-term research, various aspects of the theory of polarons
with Holstein coupling in two-level local and lattice models related with
quantum fluctuations and phase transitions  belong to
systematically studied topics in the current literature up to the
present time  \cite{Salje:1995}-\cite{Devreese:2002}.

The class of two-level electron-phonon models with linear coupling to one or two
phonon modes represent several important physical systems: excitons,
dimers, Jahn-Teller (JT) systems (rotationally symmetric E$\otimes$e and
 reflection symmetric E$\otimes(b_1+b_2)$ JT model with broken rotational
symmetry). Especially, interest in the literature is growing in the
JT models \cite{Kaplan:1995},\cite{Berlin:1997}; There is
 the experimental evidence of related structural
phase transition in some spatially anisotropic complex structures
 (perovskites, fullerides, manganites
 \cite{Berlin:1997},\cite{Gunnarson:1995},\cite{Muller:1999},
\cite{Obrien:1993}).

In two-level electron-phonon models strong enhancement of
fluctuations of phonon coordinates and momenta as well as of their
product in certain ranges of model parameters
 were reported by several authors from both numerical simulations
and analytical (e.g. variational) approaches:
Feinberg et al \cite{FCP:1990} in a model of an exciton,
Borghi et al \cite{Voit:1996} in the Holstein-Hubbard
two-level model, Morawitz et al \cite{MRK:1998} in the Holstein-Peierls
model. 

Recently,  we have investigated numerically and variationally the interplay
of quantum fluctuations and nonlinearity inherent to the two-level models
 in the ground state of the E$\otimes (b_1+b_2)$ molecular (local) JT model
  \cite{MS:2003} and the respective lattice JT model \cite{MRS:2002}.
Corresponding Hamiltonians  contain a hidden nonlinearity due to the
reflection symmetry (the nonlinearity appears explicitly under an appropriate
unitary  transformation; See Section 2 and discussion
in our recent papers \cite{MS:2003},\cite{MRS:2002}).
There is to elucidate the terminology related to quantum fluctuations and the
 nonlinearity parameters used there \cite{MS:2003} and in the present paper:
quantum fluctuations are measured by the ratio
of the phonon frequency and the classical parameter of electron-phonon
interaction, $\Omega/\alpha=1/ \sqrt{2\mu}$, while the nonlinearity parameter
is the quantum tunneling strength between the levels,
$\beta/\alpha=\chi $.
In the plane
$\mu,\chi $
($\mu=\alpha^2/2\Omega^2$)
  there occur quantum fluctuations $\sim \Omega$ and $\sim \chi$. 
The ground state in the phase plane $\mu, \chi $ exhibits
regions of dominance of either selftrapping (classical) ($\mu>1$,
$\chi<1$) or tunneling (quantum) parameters ($\mu<1$, $\chi>1$)
and regions where the classical and
quantum regions mix together ($\mu<1$, $\chi<1$ or
$\mu>1$, $\chi>1$). For $\mu<1$, formation of both the selftrapping and
tunneling "phases" is suppressed by the quantum fluctuations $\propto\Omega$.
The parameter of nonlinearity 
$\chi$ and phonon frequency
$\Omega$ are both of quantum origin so that global quantum fluctuations
mix the fluctuations due to nonlinearity close to the border of two
regions at $\chi\sim 1$ and the quantum fluctuations $\sim\Omega$.

Numerical approach to the two-level exciton and JT models in consideration in
comparison with variational approaches
bring understanding of the nature of mixing of quantum fluctuations with
the nonlinearities: since the coherent phonon subsystem does not conserve
the number of phonons the upper level participates in the distribution of
phonons even in the ground state. This manifests itself by the appearance of
additional reflective extrema of the ground state wave function
  \cite{Sander:1973},\cite{Wagner:1994}.
The distance between its extrema is related to the displacement of the
 coherent phonons and is involved into the dispersion measure of the wave
 function and of its Fourier transform \cite{MR:1997}, \cite{MM:2002},
 \cite{TR:1994}. This displacement, i.e. the measure of
 the polaron selflocalization, though suppressed by quantum
 fluctuations, is of {\it classical origin}. This is the source of
 difficulties with the justification of the moment (variance)
 characterization of fluctuations in the case of the wave function with
 the presence of the additional reflection maximum.
As it is known, in the case of multipeak wave functions,
 the variances for the coordinate and/or momentum
of related wave functions, as well as
 for other non-commuting observables do not stand for appropriate
uncertainty measures \cite{MR:1997}$^-$\cite{Buzek:1995},
since these uncertainty
measures strongly involve distance between the peaks.

Another serious shortcoming arises
when we switch from one- to two- or more phonon system, like
Jahn-Teller systems are (generally - to a system with
several non-independent variables). In this case the moment-related 
uncertainty measures are not good even for one-peak distribution, since 
the width of 
the distribution given by the variances is a width in a particular 
direction chosen arbitrarily, and cannot stand for actual ``width'' of 
a distribution.

  In these cases  the Shannon entropies of the probability density functions
assigned to such wave functions were used as alternative uncertainty 
measures of the conjugated coordinates.
The ``entropies'' of any probabilistic 
distribution are well-known to eliminate effectively both the distance between 
peaks in the multipeak distributions and the said ``anisotropy'' of the 
distribution in the space of random variables since they
contain in their expressions only functions of probability distributions,
rather than values of a random trial, as in momentum measures.

  For references, we shall briefly summarize necessary notions:
in the probability theory there are
two main types of the integral uncertainty measure assigned to an observable,
the moment and entropic ones \cite{U:1990}.
Due to existence of these uncertainty measures 
two types of uncertainty relations (UR) for two non-commuting observables
can be introduced:
(i) the Heisenberg (variance) UR and (ii)
the entropic UR. While the moment (variance) UR is expressed as the
{\it product} of the variances of two
noncommuting observables $A$ and $B$, the entropic UR is
given by the {\it sum} of their Shannon entropies.
For a continuous observable $A$ with the density of probability
distribution $p(x)$ the (differential) entropy of $A$ is defined as
 \cite{GU}
$$H_c=-\int p(x) \log p(x) dx\,.$$

The entropic uncertainty relation for the coordinate and momentum
of a quantum system described by its
normalized function $\psi (x)$ is represented by the 
inequality \cite{BBM75}, \cite{MR:1997}
\begin{equation} S_x+S_p\geq S_{xp},\label {I1}
\end{equation}
where $S_x$ and $S_p$ are the entropies of its coordinate and momentum
probability distributions
\begin{equation}
S_x= -\int\limits_{-\infty }^{\infty} \left| \psi (x) \right|^{2}
\log \left| \psi (x)\right|^{2} dx
\label{I2}
\end{equation}
and
\begin{equation}
S_p=-\int\limits_{-\infty}^{\infty}\left| \varphi (p)\right|^{2}
\log \left| \varphi (p)\right|^{2} dp,
\label{I3}
\end{equation}
respectively, and $ \varphi(p)$ is the Fourier
transform of the wave function  $\psi(x)$  and $S_{xp}$ represents
the lower bound of the right-hand side of the inequality (\ref{I1}).

It has been found by Bia\l ynicki-Birula and Mycielski \cite{BBM75}
that the lower bound $S_{xp}=1+\log \pi $ for the harmonic oscillator
represents the minimal lower  bound for any quantum system with one pair of
non-commuting observables.
Therefore, the sum of coordinate and
 momentum entropies of the arbitrary quantum system is $(\hbar =1)$
\begin{equation}
S_x +S_p\geq 1 +\log \pi.
\label{I4}
\end{equation}
This relation is an entropic counterpart of the famous Heisenberg principle 
formulated by means of variances: $\Delta x \Delta p \ge 1/2$.

 We note that the entropy as a measure of uncertainty and the
entropic uncertainty relations are widely used in quantum optics
  \cite{Buzek:1995},\cite{Ohya:1993} being useful
 for systems with more complex structure of photon coherent states
 (e.g., Schr\"odinger cat coherent states).

In Section 2, two standard two-level electron-phonon models with
linear electron-phonon coupling are described as prototype models for
which the entropic uncertainty principle represents the adequate
measure of the fluctuations. The nonlinearity
hidden in the (initially linear) two-level phonon Hamiltonians is revealed
after appropriate diagonalization of the problem with respect to the
electronic subspace.

In Section 3 we determine by direct numerical simulations the wave
functions of one-phonon exciton model
for a range of model parameters.
From those, we determine and compare the quantum fluctuations as
functions of the effective interaction $\mu $
described by means of the Shannon entropies of the phonon coordinate and
momentum on one hand and by means of their variances on the other hand.
Comparison of the sum of the coordinate and momentum Shannon entropies
and the corresponding product of the variances
clearly puts forward the former as more adequate measure of the quantum
fluctuations in certain parameter region than the latter one.

Our approach is consequently based on numerical simulations,
because variational treatments based on squeezed coherent states or
their linear combinations were generally found to underestimate
quantum fluctuations
: (i) they underestimate quantum fluctuations originating from finite phonon
frequency \cite{FH:1982},\cite{ZFA:1989} for weak couplings;
(ii) The fluctuations due to the
nonlinearity  in the crossover region are strongly
coupled with the fluctuations (i). Therefore,
in the region close to the crossover there appear discontinuities which are
artefacts of variational approaches (similar to those of the adiabatic
approximation).

In Subsection 3.1 leaning upon
variational treatment of the exciton model we present results of
simple analytical estimations of quantum fluctuations (both entropic 
uncertainties and 
variances) 
to illustrate
the failure of variational methods. However, in a small range of the
phase plane we can identify the effect of the classical displacement
in the Heisenberg variances and their product by means of variational
method.

Similar investigations for the 
two-mode phonon $E\otimes(b_1^{\dag}+b_2)$ Jahn-Teller model
(which differs from the
model of Section 3 by the presence of phonon assistance in the
tunneling term and, consequently, the existence of
additional mode correlations effects) 
are presented in Section 4.

 \section{Model Hamiltonians}

 For the model of interest let us consider the two-level reflection
symmetric electron-phonon system with linear (Holstein) coupling
to one or two phonon modes
\begin{equation}
H= \sum _{i=1}^{N=1,2}\Omega(b_i^{\dag}b_i +1/2) +
\alpha (b_1^{\dag}+b_1)\sigma_z- \beta \Lambda \sigma_{x},
\label{1}
\end{equation}
where the Pauli matrices $\sigma_x, \sigma_z$ represent electron
density operators in the pseudospin notation,
$\sigma_z=\frac{1}{2}(c_2^{\dag}c_2-c_1^{\dag}c_1),
\sigma_x=\frac{1}{2}(c_1^{\dag}c_2+c_2^{\dag}c_1)  $ with
$I=c_1^{\dag}c_1+c_2^{\dag}c_2 $ as the unit operator
 \cite{MRS:2002}.

For the case of coupling to one phonon mode, $N=1$, we take
$\Lambda=1$ and two-level Hamiltonian (\ref{1}) represents an exciton
or a dimer with tunneling between the levels with the
tunneling amplitude $\beta$. 
The prototype model has been handled both variationally and numerically
by many authors in its local and extended version
 \cite{FCP:1990},\cite{Sander:1973},\cite{Wagner:1994},
\cite{Herfort:2001}.
In the case of the
coupling with two phonon modes, $N=2$, the second term with
$\Lambda = b_2^{\dag}+b_2$ represents phonon-assisted tunneling between the
levels (the flip-flop rate is proportional to the value of phonon-$2$ 
coordinate, i.e. an additional ``phonon pumping'' occurs in distinction to 
the one-phonon case).
The special case $\alpha=\beta$ represents
standard $E\otimes e $ Jahn-Teller Hamiltonian with rotational symmetry
 \cite{Barentzen:1977},\cite{Lo:1991},\cite{Wagner:1992},
\cite{Obrien:1993},\cite{Barentzen:2001}.
The phonon $1$-mode is antisymmetric and the $2$-mode is symmetric
against the reflection.
The case $\alpha\neq \beta$ is the reflection symmetric
$E\otimes (b_1+b_2)$ Jahn-Teller Hamiltonian
 recently investigated variationally and 
numerically \cite{MS:2003},\cite{MRS:2002}.

 The reflection symmetry of
the Hamiltonians (\ref{1}) involves a nonlinearity which reveals itself
explicitly after performing Fulton-Gouterman unitary
transformation  (Fulton et al \cite{Fulton:1961}, Shore et al \cite{Sander:1973}):

\begin{eqnarray}
H_{FG}=U H U^{-1} = \Omega\sum_{i=1}^{N=1,2}(b_i^{\dag}b_i +1/2) +
\alpha (b_1^{\dag}+b_1)I \nonumber \\ 
\mp \beta \Lambda \exp (i\pi b_1^{\dag}b_1),
\label{2}
\end{eqnarray}
where $U= \frac{1}{\sqrt 2}\left ( \matrix{1,\ G\cr 1, \ -G}\right ),
\quad G= \exp (i\pi b_1^{\dag}b_1)$ is the reflection operator in the phonon space,
$G (b_1^{\dag}+b_1)=-(b_1^{\dag}+b_1) G$,  which performs virtual coupling 
of the levels by
phonons $1$  mediating the electron (Rabi) oscillations between them.

The transformation (\ref{2}) diagonalizes (\ref{1}) in the electronic subspace
yielding however a strong nonlinearity in the phonon subspace (term
containing $\beta$) which otherwise was hidden in the initial 
Hamiltonian (\ref{1}). The unitary transformation left us with purely
phonon Hamiltonian  (\ref{2}) while electrons were excluded. All further 
consideration will be performed for the ground
state which for both models 
is given by the upper sign in the transformed Hamiltonian (\ref{2}).

The competition of quantum fluctuations ($\Omega$), the selflocalization
($\alpha$) and the phonon assisted tunneling ($\beta$)  in (\ref{2})
determines regions of dominance of said effects for different sets of
the parameters. In the next sections, we shall analyze these regions by
numerical simulations of the models (\ref{2}).

\section{Exciton: coupling to a single phonon mode}

The main purpose of this and of the next  section is to show that
for the considered prototype models the
entropic uncertainties and the entropic uncertainty principle represent
adequate measures of quantum fluctuations instead of the moment
(variance) Heisenberg ones.

The exciton model, including respective wave functions, has been previously
intensively investigated by many authors; Here we shall confine ourselves
to essential points needed for our purposes.

The results of the numerical simulations of the ground state wave functions
of the model (\ref{2}) for $\Lambda=1$, i.e. of one-mode diagonal coupling
case, are shown in Fig. 1.
The parameter of the effective interaction $\mu=\alpha^2/2\Omega^2 $
is the measure of the competition between the classical polaron
selflocalization due to the interaction of the energy $
\alpha^2/2\Omega$ (e.g., Holstein \cite{Holstein:1959})
and of the quantum fluctuations of the energy $\Omega$.
Parameter $\beta/\Omega $ is the measure of the competition between the
quantum term of the tunneling $\beta$ and quantum fluctuations $\Omega$.
Wave functions shown in Figs. 1 illustrate the effects of the interplay
between these parameters affecting the population of the phonons
on both levels.
One can see that the most prominent deviations from the one-peak wave
function occur for large $\mu$; For moderate $\mu$ and $\beta, \
(\Omega=1)$ the wavefunctions also tend to exhibit variations from
almost-Gaussian shapes.
These features are consistent with non-conservation of the number of the
coherent phonons involved. They inspired the variational approach based
on linear combination via a variational parameter of two harmonic
oscillators related to both levels \cite{Sander:1973},\cite{Wagner:1994}.
Numerical evaluation of the ground state (Fig. 2) illustrates
smooth decrease of the energy as function of $\mu$ indicating
 a smeared crossover region.
The sharp transition line between two regions resulting from the
competition of two interactions ($\mu$ and $\beta$)
is known to be an artefact of the adiabatic approximation and often of
inadequate variational approaches.
In the quantum treatment the crossover line is smoothed
over the region of a width $\approx\Omega$ of the phonon energy.

These regions are analogous to
the "selfrapping dominated" and the "tunneling dominated" regions for the
$E\otimes (b_1+b_2)$ model \cite{MS:2003},\cite{MRS:2002} (see the 
next Section).
In the weak coupling region, $\mu<1 $, the
formation of the "ordered phases" (either selftrapping- or tunneling-
dominated) is suppressed by the quantum fluctuations $\sim\Omega$ as
well as in the model of Section 4.

Corresponding to the above described reasons for qualifying the quantum
deviation from the harmonic oscillator (Gaussian) behaviour
 we evaluated numerically (Fig. 3a) the left-hand side of the Shannon uncertainty
 relation, the sum of entropies $S_{Q}+S_{P}$ (\ref{I4})
(here, $Q$ and $P$ are the  coordinate and momentum of the phonon).
The product of variances  $\sqrt {\Delta Q^2 \Delta P^2}$
containing explicitly the displacement (distance between the peaks)
is shown on Fig. 3b.
Both approaches, the variance and the Shannon entropy ones, can be
compared to illustrate their different physical content.
Close to the line of equal effective coupling and the tunneling
strength  $2\mu=\beta $ the extremum of the sum of Shannon entropies
 manifests itself markedly. The pictures of related wave functions
attribute this extremum by the maximal overlapping (tunneling) of
the contributions of both levels (Fig. 1b).
With growing polaron selflocalization $\mu$ the quantum fluctuations are
suppressed and both oscillator parts tend to separate, see Fig. 3a for
large $\mu$.
The harmonic oscillator value $1+\log\pi= 2.14473$ is recovered in 
both regions around the crossover region: in the
limits of large  $\mu$ or, on the contrary, for large $\beta$ and small 
or moderate $\mu$ (less than some critical value on the crossover line).

The product of variances $\sqrt {\Delta Q^2 \Delta P^2 }$
at Fig. 3b shows similar behaviour except for large $\mu$,
where it therefore accounts for the growing distance of the peaks of the wave
functions with $\mu$, Fig. 1d.
This distance as a function of a {\it classical parameter} $\mu$ is a
measure of the nonlinearity since it maps the position of the reflection
peak. In the limit of small $\mu$,  $\sqrt {\Delta Q^2 \Delta P^2 }$ tends to
the harmonic oscillator value 0.5.

The range of maximum fluctuations in the plane $\mu, \beta$ (Fig. 3)
 separates the quantum fluctuation dominated region of
the harmonic oscillator $\Omega$ ($2\mu < \beta$) and the selftrapping
dominated region ($2\mu> \beta$). Although both uncertainty
measures yield similar behaviour in the crossover range  showing the common
increase of fluctuations, and in the tunnelling region where
the wavefunctions indeed converge to an oscillator,
the quantitative information beyond 
the crossover region is different in the selftrapping region.
Enhancement of fluctuations reported by the variance uncertainty picture
in the selftrapping region point
merely on the {\it classical} contribution related to the displacement
so that {\it it fails
to play a role of a measure of quantum fluctuations}. This shortcoming
is effectively eliminated by the Shannon entropic relations which thus
help to extract purely quantum effects manifested in our case in the strongly 
non-Gaussian behaviour of single peaks.

Figs. 3c, d compare crossections of both Figs. 3a, b as functions of
$\mu$ for different $\beta$. The entropies and variances plotted on one graph
(both shifted by the oscillatory values of corresponding quantities)
illustrate the difference of two approaches:
the Heisenberg product of variances with increasing  $\mu$ (upper curves 
in Fig. 3c,d) is
growing while the sum of entropies tends to the oscillator value as
expected. 
(For quantitative comparison we should plot rather
exponents of the sum of entropies than the entropies; this becomes clear 
if we remind of the famous
Einstein relation between entropies and fluctuations $\exp(S)\sim \Delta X^2$,
but the difference between them is almost inappreciable).

 In the next Subsection we shall
illustrate these considerations by means of analytical calculations of
both measures based  upon a variational fitting of wavefunctions.

The enhancement of quantum fluctuations for the exciton model have been
found by Feinberg, et al. \cite{FCP:1990} by numerical calculation
of the moment (variance) uncertainties and their product as functions of
parameters $\alpha^2\equiv 2\mu $ and $\lambda\equiv 2\mu/\beta$. For given
$\alpha$, their dependence on $\lambda$ means in our notations 
the dependence on $1/\beta$.
Let us note that these fluctuations are related to the competition of the
parameters $\mu $ and of the nonlinearity parameter $\beta$,
i.e. related to a crossover between the
selftrapping dominated and the tunneling dominated regions.
However, the undesirable
presence of the classical nonlinearity manifests itself in the
dependence of $\sqrt{\Delta Q^2\Delta P^2}$ on $\mu$ due to the
competition between the parameters $\alpha $ and $\Omega$.
Namely, it becomes
obvious when the product of variances is depicted as a function of
$\alpha^2\equiv 2\mu$ as in Fig 3.

\subsection{Analytical illustration: variational approach}

Though the variational approaches based on squeezed coherent states are
not suitable for description of quantum fluctuations
we will apply them
for evaluations of fluctuations of a simple exciton model. We can use them
for demonstration of the effect of interest (i.e. of the involvement of the
displacement in the moment uncertainties product in contrast to its entropic
counterpart). We will  use variational
approach with two squeezed coherent harmonic oscillators linearly combined
by a variational parameter (\ref{VA2}) in accordance with the concept
initiated by
Shore and Sander \cite{Sander:1973}. However, this illustrative analytics is
valid only if the maxima of the wave function are well separated from each
other (small overlapping), i.e. for strong coupling (large $\mu$). This
is just the region of the most prominent difference of both said concepts of
the uncertainty measure.

Phonon variational wave function for the two-level model with
one phonon mode ($\Lambda=1$) (\ref{2}) as solution to
the Fulton-Gouterman equation
 \cite{Wagner:1994},\cite{Fulton:1961},\cite{Herfort:2001}
\begin{equation}
H_{FG}^{(p)} \phi ^{(p)}  =
[\Omega\left (b^{\dag}b+1/2\right ) +\alpha (b^{\dag}+b)  - p
\beta G ) ]\phi^{(p)} =E^{(p)}\phi^{(p)},  
\label{5}
\end{equation}
$p= \pm 1$

for the ground state ($p=1$) can be chosen in the form
 \cite{Sander:1973},\cite{Wagner:1994}:
\begin{equation}
|\Psi ^{(1)}\rangle=  \frac{1}{\sqrt  C}(1+\eta  G)|\phi^{(1)}\rangle ,
\label{6}
\end{equation}
where $G=\exp(i\pi b^{\dag}b)$ is reflection operator in the phonon
space, $G |\phi^{(1)}\rangle= |\phi^{(2)}\rangle  G$, and
\begin{equation}
|\phi^{(1,2)}(\gamma, r)\rangle =D(\pm \gamma)S(r)|0\rangle .
\label{VA1}
\end{equation}
The indices $1 (2)$ pertain to the lower (upper) level.
The functions $D(\gamma)$ and $S(r)$ are generators of displacement
and squeezing depending on the variational parameters:
\begin{eqnarray}
D (\gamma)= \exp [\gamma (b^{\dag}-b)],\nonumber \\
S(r)= \exp[r(b^{\dag 2}-b^{2})]
\label{VAA}
\end{eqnarray}
acting on the phonon vacuum $|0\rangle$.
The normalization constant in (\ref{6}) is $C= 1+\eta^2+2\eta\exp
[-2\gamma^2 \exp{(-4r)}]$.
Wave function (\ref{6})-(\ref{VAA}) represents two squeezed displaced
harmonic oscillators combined by a variational parameter $\eta$ 
(compare with Fig.1).

Variational parameters of the displacement $\gamma$, squeezing $r$ and
the parameter of admixture of the reflection part of the upper level $\eta$
can be determined by minimalization of the Hamiltonian (\ref{5}) averaged over 
(\ref{6}):

\begin{eqnarray}
\langle H \rangle = \frac{1}{2}\cosh{4 r}+ \nonumber \\
\frac{1}{C}\gamma^2\left(
1+\eta^2-2\eta \exp{(-8 r)}\exp{(-2 \tilde{\gamma}^2)} \right) +
\nonumber \\
\frac{2 \alpha}{C} \left( 1-\eta^2 \right) \gamma 
 - \frac{\beta}{C} \left( (1+\eta^2)\exp{(-2\tilde{\gamma}^2)}+2\eta
\right), \nonumber \\
\quad\tilde\gamma\equiv \gamma \exp(-2r).
\label{Ham:trial}
\end{eqnarray}
(The parameters are scaled so that $\Omega=1$).

It is worth noting that
variational approaches of varying degree of reliability combined with
unitary transformations are widely used for electron-phonon
systems
 \cite{FCP:1990}$-$\cite{Wagner:1994},\ \cite{Herfort:2001},
\cite{Lo:1991},\cite{ZFA:1989},\cite{Wagner:1986}.
Comparison of different variational Ansatzes for one-phonon two-level
system was performed e.g. by Shore et al \cite{Sander:1973},
Sonnek et al \cite{Wagner:1994} and for
E$\otimes(b_1+b_2)$ model by the present authors \cite{MS:2003}.

Let us evaluate variances of the phonon
coordinate $Q= (b^{\dag}+b)/\sqrt 2$ and momentum $P=
(b^{\dag}-b)/i\sqrt 2 $,   $(\Delta Q)^2= \langle Q^2\rangle-\langle
Q\rangle^2,  $ $(\Delta P)^2= \langle P^2\rangle-\langle P
\rangle ^2,  $ where we average over the states (\ref{6})-(\ref{VAA}).

We get
\begin{equation}
(\Delta Q)^2= \frac{1}{2}\exp(4r) \left ( 1+ 8\eta\tilde
\gamma^2\frac{2\eta+(1+\eta^2)\exp(-2\tilde\gamma^2)}{(1+\eta^2+2\eta
\exp (-2\tilde\gamma^2))^2}\right),
\label{VA2}
\end{equation}
\begin{equation}
(\Delta P)^2= \frac{1}{2} \exp(-4r)\left ( 1-8\eta\tilde
\gamma^2\frac{ \exp(-2\tilde\gamma^2)}
{1+\eta^2 +2\eta \exp (-2\tilde\gamma^2) }\right) .
\label{VA3}
\end{equation}

For large $\gamma$, the product of variances can be estimated by
\begin{equation}
(\Delta Q)^2 (\Delta P)^2 \simeq \frac{1}{4}\left
(1+\frac{16\eta^2}{(1+\eta^2)^2}\tilde\gamma ^2\right ),
\label{VA4}
\end{equation}

where $\tilde \gamma$ is defined by (\ref{Ham:trial}).
 From (\ref{VA2}), (\ref{VA3}) and (\ref{VA4}) it is obvious that the anomalous
 enhancement of fluctuations is due to the contribution of the
classical displacement (reduced by squeezing) $\tilde\gamma$ and is mediated by the
reflection parameter $\eta$. If $\eta=0$, we are left with a single harmonic
oscillator as expected. 

It is easy to perform simple illustrative analytical estimations for
the values of variational parameters as functions of the model parameters. 
Assuming in (\ref{Ham:trial}) $r, \eta$
small,  we get an approximate equation for $\gamma$ setting $\partial
\langle H \rangle /\partial \gamma=0$
(compare similar procedure in the recent paper \cite{MS:2003}),
\begin{equation}
\gamma\left( 1+2\beta\exp{(-2\tilde \gamma^2)} \right)=-\alpha\, ,
\label{gamma:est}
\end{equation}
(from similar considerations the expression for $\eta$ can also be
found). 

Fig. 4 illustrates results of the variational approach for the Heisenberg
product of variances: in comparison with Fig. 3b, there occurs a
complete suppression of fluctuations in
the tunneling region $\sim 2 \mu <\beta$. Instead of the smooth crossover
there occurs sharp discontinuity which is an artefact because of
underestimation of the fluctuations $\sim\Omega$ similar to that of
adiabatic approximation. In the selftrapping region,  $\sim 2\mu>\beta $,
the increase is evidently caused by the classical contribution
because quantum fluctuations tend to disappear in the classical limit of
the strong coupling. 

To illustrate the advantages of the entropic uncertainty relation over the
moment Heisenberg ones for our case we calculate approximately the
expressions for the Shannon entropies of coordinate and momentum for the ansatz
(\ref{6}), (\ref{VA1}). This can be easily done analytically if $\eta \ll 1$
and $\gamma \gg  \exp(2 r)$ (last condition meaning that two peaks of the 
wavefunction are well separated and almost do not overlap).

For the entropies $S_Q$ and $S_P $ (\ref{I2})-(\ref{I3}) with
(\ref{6})-(\ref{VAA}) we get
\begin{equation}
S_Q= \frac{1}{2}(1+\log \pi)+\log (1+\eta^2)+2r
-\frac{\eta^2}{1+\eta^2}\log (\eta^2) +O(\varepsilon,\eta^3),
\label{VA$}
\end{equation}
\begin{equation}
S_P= \frac{1}{2}(1+\log \pi) -
2r-\frac{\eta^2}{(1+\eta^2)^2}+ O(\varepsilon,\eta^3), \quad
\varepsilon\equiv \exp (-2\tilde\gamma^2).
\label{VA5}
\end{equation}
From (\ref{VA$}) and (\ref{VA5})
\begin{eqnarray}
S_Q+S_P= 1+\log \pi +\log(1+\eta^2) -\frac{\eta^2}{(1+\eta^2)^2}
-  \nonumber \\
- \frac{\eta^2}{1+\eta^2}\log (\eta^2) +O(\varepsilon,\eta^3).
\label{VA6}
\end{eqnarray}
 For $\eta=0$, Eq. (\ref{VA6}) reduces to the single oscillator value
 $1+\log \pi$ as expected.
Contrasting (\ref{VA6}) with (\ref{VA4}) we see that entropy
uncertainty relations are weakly dependent of the displacement $\tilde\gamma$
and contain as the main contribution owing to the parameter $\eta$ originating
from the nonlinear effect due to coupling between levels.

\section{Interplay between quantum fluctuations and
nonlinearity in $E\otimes (b_1+b_2)$ Jahn-Teller model.}

Quantum ground state of $E\otimes (b_1+b_2)$ Jahn-Teller
reflection symmetric model (a degenerate electron
level coupled with two phonon modes, one symmetric and one
antisymmetric against the reflection) has been investigated in our recent
papers  for the one site \cite{MS:2003} and lattice \cite{MRS:2002} case.
Let us expound the main results relevant to
clarify the origin of quantum fluctuations of interest.

Exact numerical simulations of the solution to the Hamiltonian (\ref{2})
with $ \Lambda = b_2^{\dag}+b_2  $ yield the ground state wave functions
depicted in Figs. 5 a, b, c (in the coordinate representation in the space
$(Q_1\otimes Q_2)$), in terms of the parameters $\mu=\alpha^2/2\Omega^2$ and
$ \chi=\beta/\alpha $ for three characteristic regions \cite{MS:2003}.
The two-phonon ground state wave functions for the case of
phonon-assisted tunneling exhibit evident mixing of the
phonon-1 wave functions related to the lower and excited levels at
$\chi<1 $, Fig. 5a.
Each of the levels refers to one of two competing minima of the effective
 potential composed of $\alpha $- and $\beta$-components, as
described below, Fig. 6.
    The mixing due to the two mode correlation 
is pronounced most effectively in the region of
the dominant quantum fluctuations $\Omega$, $\mu <1, \chi\simeq 1 $
 \cite{MS:2003}.
At $\chi>1$, the prominent peak refers to a harmonic oscillator
of the dominant potential well $\beta$ related to the selflocalized state
of  an electron oscillating between the levels (Fig. 5c).
Because of the phonon-$2$ assistance there occurs a smeared continuous
crossover from the regime of selflocalization (Fig. 5a) towards the
tunnelling regime (Fig. 5c) through the intermediate picture close to
the E$\otimes$e Jahn-Teller case (Fig. 5b).

In spite of the limitations of variational approaches (their failure close
to the crossover between two regimes,
underestimation of quantum fluctuations) they provide
useful insight into the behaviour of the ground state.
The complex interplay
of the nonlinear and quantum effects has been analyzed by numerical
simulations and compared with results of various variational
treatments \cite{MS:2003}.
 We note that the variational wave functions with the
admixture of the excited symmetric phonon mode (phonons-$2$) we 
proposed recently \cite{MS:2003} for
 the ground state showed significant
 improvement of the agreement  with numerical simulation results
 especially for strong  e-ph couplings $\mu$.
  The topology of the effective variational potential, Fig. 6,
 (Hamiltonian (\ref{2}) averaged over trial functions depending on a set of
variational parameters) is controlled by several model parameters;
it plausibly can acquire two or more competing minima
referring to the ground state with a possible admixture of the
side minimum referring to the excited state \cite{MS:2003}. As a result,
two regions of the ground state appear according to which one of
two local minima of the potential dominates;
generically two regions are recognized - with either dominating  
selftrapping ($\chi<1$) or tunneling ($\chi>1$). The existence of 
selftrapping dominated
vs tunneling dominated regions results from complex competition of two
nonlinearly coupled coherent phonon modes ($\beta$-term in (\ref{2})).
Let us note, that the order parameters for the selftrapping dominated
"phase" is the displacement $\langle b_1^{\dag}+b_1\rangle\sim Q_1$ and for the
tunneling dominated "phase" $\langle b_2^{\dag}+b_2\rangle\sim Q_2$.
These
features are analogous for both local \cite{MS:2003} and a generalized
lattice \cite{MRS:2002} E$\otimes(b_1+b_2)$ JT model.

The additional phonon-2 assistance of the tunneling with
non-conservation of the number of the phonons 2 implies their selfconsistent
behaviour and  creation of a corresponding potential well which
develops proportionally to the interaction strength $\beta$ (Fig. 6).
At the same time the correlation of both phonon modes is involved
(last term of (\ref{2})) via the multiple (Rabi) tunneling mediated by
the mode 1 as a source of the nonlinearity.
 The nonlinearity significantly enhances the quantum fluctuations
 in the region of maximal tunneling  between the levels \cite{MS:2003}.
 Competition of the terms responsible for selftrapping and
 phonon-assisted tunneling between the levels resulting in formation of
 two ground state regions is accompanied by increased anomalous
 quantum fluctuations in the crossover region. There is to be
 emphasized that, similarly to the case of $\Lambda=1$ (Fig. 2),
 there exists no sharp transition line in the ground state of the energy
 in the phase diagram.
 The transition region is smeared by a width of the phonon frequency $\Omega$.
Sharp transition line occurs rather as a well known artefact of
some  variational approaches and of the adiabatic
approximation \cite{Sander:1973}.

In order to describe the quantum fluctuations resulting from the complex
interplay of the above described contributions we have numerically
evaluated the Shannon entropies $S_{Q}$, $S_{P}$ (Figs. 7a, b)
and their sum $S_{Q}+S_{P}$ (Fig. 8a) as functions of $\mu$ and $\chi$.
Here $Q\equiv \{Q_1,Q_2 \}$ and $P\equiv \{ P_1, P_2 \}$ and all integrations 
are meant in the two-dimensional space $Q_1 \times Q_2$, resp. $P_1 \times P_2$.

In the weak coupling region, $\mu=\alpha^2/2\Omega^2 \leq 1$,
formation of the "phases" at $\chi<1$ and $\chi>1$ is greatly
reduced because of the fluctuations $\Omega $ and
strong correlations between the phonon modes (Fig. 8a). For large
$\mu$ the extrema
at $\chi\simeq 1$ get sharper but remain non-singular due to the finite (although
small) quantum fluctuations $\Omega$.
 The obvious asymmetry of the sum of entropies $S_{Q}+S_{P}$
as a function of $\chi$ comes from the fluctuations of the $\chi>1$
"phase". In the limit of large $\chi$, both uncertainties, entropic
and momentum ones, tend to their values corresponding to two harmonic
oscillators,  $2(1+\log \pi)=4.28946$ and $0.5$, respectively
(Figs. 8a, b or 9a, c).
This resembles much the tunneling dominated
region of one-phonon model (Section 3) with the single exception
that the tunneling
region is spread for all $\mu$ because of the pronounced crossover
enhanced by phonon-$2$ pumping.

Numerical result for the product $\sqrt{\Delta Q_1^2\Delta P_1^2}$
as function of $\mu$ and $\chi$ is shown in Fig. 8b and 9b, d.
The contour plots
of Figs. 8 presented on Fig. 10 help visualizing their difference,
namely
(i) the classical contribution of the displacement growing with the coupling
$\mu$ to the Heisenberg product, Fig. 10. b, and
(ii) prevention of formation of the selftrapping dominated and the
tunneling dominated regions in the weak coupling region $\mu<1$. Except for
the region $\mu \ge 2$ the difference between Fig. 8a and 8b is seen not so clear 
as on the corresponding Fig. 3 a),b) for one-phonon model. However it can be easily 
traced if one compares the corresponding crossections along $\chi$ and $\mu$ axes, 
visualised on Fig. 9a)-d). In particular, of main interest for the application of 
suggested alternative measure are regions of large $\mu$ close to $\chi\simeq 1$ and
of $\chi \le 1$ where the wavefunctions exhibit pronounced multipeak structure, 
as seen from Fig. 5a),b).

The region of maximum in Figs. 8a, b at $\mu\simeq 1$ refers to the
crossover between the quantum
fluctuation dominated region $\mu<1$ and the selflocalization dominated
region $\mu>1$.
Namely, at $\mu\sim 1$ there occurs a crossover from
two correlated oscillators towards two independent oscillators, which is
visualized in Fig. 8a.
These two independent oscillators (see wavefunctions on Fig. 5a) refer to the
complexity of the  classical (i.e. adiabatic limit) potential and the 
entropic relations handle them accurately marking this region as
a ``classical'' one with close to two-oscillator value $2(1+\log \pi)$.
  On the contrary, the variances in the left-hand side
  of the Heisenberg uncertainty relation (Fig. 4b) continue increasing
for small
$\chi$ with growing $\mu$, but apparently weaker than in the one-phonon
case of Section 3.
In the two-phonon case the separation of two peaks of the wave function
in the selftrapping region is much less pronounced than in the one-phonon
case, which is seen
also from the moderate increase of the product of variances at Fig. 8b,
9c, 10b, for
$\chi<1$ in comparison with the corresponding one-phonon case at Fig. 3b.

\section{Conclusion}

Two-level electron-phonon systems with one (exciton) and two (E$\otimes
(b_1+b_2)$ Jahn-Teller model) phonon modes exhibit reflection symmetry
as the source of a hidden nonlinearity which reveals explicitly
by appropriate unitary diagonalization (Fulton-Gouterman transformation (\ref{2}))
of the Hamiltonian.
This diagonalization (i) excludes
electrons leaving us with solely phonon Hamiltonian (exact decoupling in 
electron subspace), 
(ii) reveals 
quantum correlation of phonon modes in the case of two-phonon model.
As a consequence, related phonon wave functions exhibit
multipeak, i.e. essentially non-Gaussian structure even in the  ground
state. Besides this "topological" anisotropy, additional anisotropy
appears in two-phonon model due to the correlation of the modes. 
Resulting complex topology
of the ground state implies appearance of two regions
in the phase plane $\mu, \chi$ or $\mu,\ \beta$: the
selftrapping  dominated region for $\chi < 1$ or $2\mu>\beta$ and the
 tunneling dominated region for $\chi >1$ or $2\mu<\beta$ respectively,
for both models due to
competition of two antagonistic (classical selftrapping and quantum
tunneling \cite{MRS:2002},\cite{MS:2003}) interactions in the 
Hamiltonian (terms with $\alpha$ and $\beta$
in (\ref{1})).  The enhancement of fluctuations close to the crossover
between these two regions 
behaves in analogy with
the corresponding items  of the theory of critical phenomena but appears 
smoothed by
the finite phonon frequency $\Omega$.
Namely, quantum  fluctuations due to finite phonon frequency
$\Omega$ prevent formation
of the ordered "phases" in the region of weak coupling $\mu<1$ in the
phase plane $\mu, \beta$ or $\mu, \chi$. The interplay of the
classical and quantum terms yields complex interplay (mixing) of
quantum fluctuations $\propto\Omega$ and nonlinear fluctuations $\propto\chi$.

The moment and entropic uncertainty measures represent two classes of
characteristics used in the probability theory in order to describe 
quantitatively the
spreading of the probability distribution for an observable. The uncertainty
principle in quantum mechanics leans essentially on the probabilistic 
interpretation of the wavefunction and can be thus formulated in a
twofold fashion - in the form of either moment (variance) or entropic
uncertainty relation.

The aim of this paper was to show that for certain parameter region 
the appropriate quantitative measure of
phonon quantum fluctuations of the systems in consideration
could be the
entropic uncertainty measures rather than the moment measures (variances)
 used in the Heisenberg uncertainty relation.
We have compared results of numerical
simulations for both types of measures and show that the latter exhibits
serious shortcomings in description of the global fluctuations of such
systems.

Namely,  Heisenberg uncertainty relations
impose the measure under integration ($Q_i^2$, resp. $P_i^2$) which is
noninvariant in the space of variables - both translationally and rotationally.
Hence there are two sources of the disqualification of the Heisenberg
 variances,
both resulting from the anisotropy of wavefunctions caused by
the reflection symmetry  of considered models:
(i) former non-invariance has its consequences in difficulties describing the 
multipeak distributions, and (ii) latter one - when trying to judge about two or
more strongly correlated variables.

Problem of characterization of quantum fluctuations by variances for a
system of two correlated oscillators is much more subtle than for
one oscillator of Section 3. In addition to the problems
imposed by multipeak distributions a serious difficulty arises if
several coupled variables are involved (in our case - two strongly correlated
phonon oscillators). If for one-mode case (and for the probability
distribution with a single peak) $\langle\Delta X^2 \rangle$ could stand for 
an effective ``width'' of the distribution function, in many-variable case 
it represents only an average ``width'' in the $X$-direction, that is the 
measure under the integral is not the best choice since it is strongly 
affected by, e.g., basis rotation. As an example just consider a squeezed 
correlated two-phonon trial function \cite{MS:2003} of the general form
$\exp{(-\sum_{i,j=1,2}a_{ij}x_ix_j})$
representing an arbitrarily turned ellipsoid in $(x_1,x_2)-$plane.
To judge about the ``width'' of such distribution in terms of variances 
one should therefore consider rather tensor quantities representing the 
``width'' of the distribution for every direction \cite{descr}.

The entropic measure, on the contrary,
does not suffer from this shortcoming.
It presents a handy characteristics of
essentially scalar character which can be applied for any number of coupled 
variables, but is not affected by the basis rotation.
This holds generally for a measure which suggests taking an average 
of some function of the distribution itself (averaging of $\log P(x)$, as for
Shannon entropy considered here, or generalized entropic measures, like, for 
example, R\'{e}nyi entropy \cite{MMS:2003}).

Similar problems arise, e. g. in quantum
optics, at the description of fluctuations of the photon
coherent Schr\"o\-din\-ger cat states and photon multimode correlated
systems (see, e.g. Ohya et al \cite{Ohya:1993} and Bu\v zek \cite{Buzek:1995}
and references therein). The entropic uncertainty relations are commonly
recognized there as a good tool for handling fluctuations of quantum origin.

\vspace{1cm}

{\small Acknowledgements

The support from the Grant Agency of the Czech Republic of our
project No. 202/01/1450 is highly acknowledged.
 We thank also the grant agency VEGA (No. 1/0251/03) for partial
 support.}



\onecolumn

\begin{figure}[h]
\includegraphics[scale=0.9]{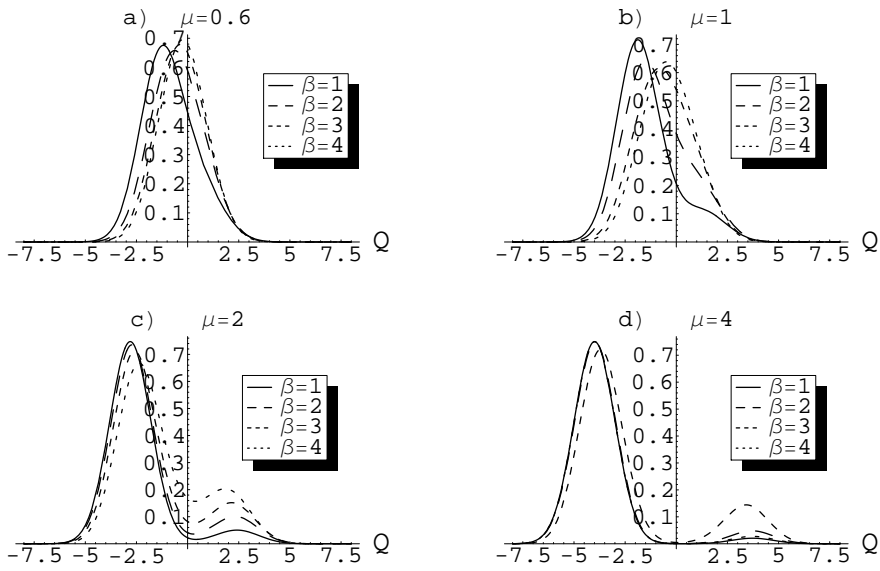}
\caption{Ground state wave functions of one-phonon model parametrized by
$\beta$ and $\mu$, $\mu=0.6$ a); $\mu=1$ b); $\mu= 2$ c); $\mu=4$ d);
$\Omega=1$.}
\label{fig1}
\end{figure}


\begin{figure}[h]
\includegraphics[scale=0.7]{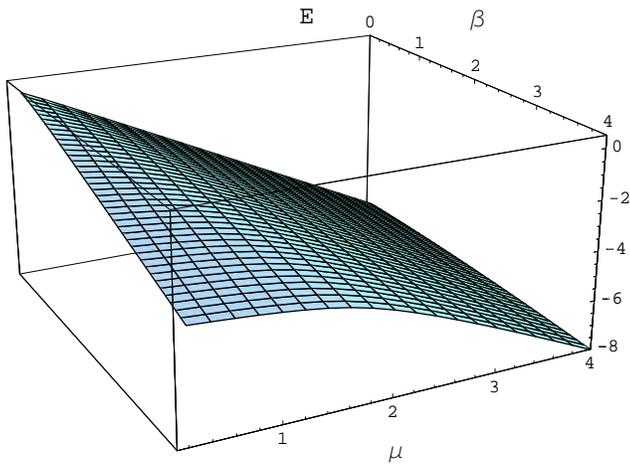}
\caption{Ground state energy in the plane $\mu$ and $\beta$.}
\label{fig2}
\end{figure}


\begin{figure}[h]
\includegraphics[scale=1.05]{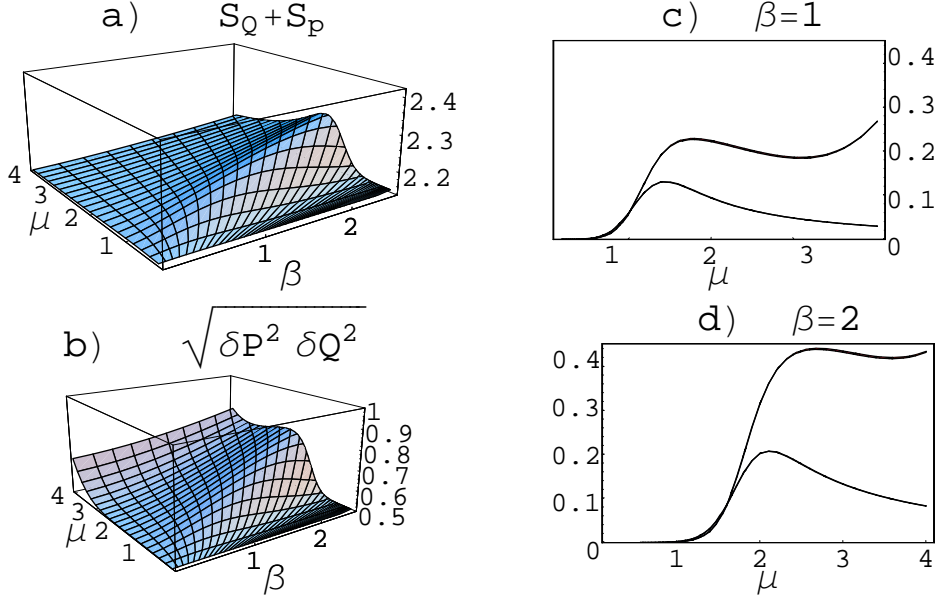}
\caption{Sum of entropies $S_{Q}+S_{P}$ (a) vs product of variances
$\sqrt {\Delta Q^2 \Delta P^2}$ (b) for one-phonon model; $\Omega=1$.
  Product of variances (upper curves)
and exponent of the sum of entropies (lower curves) shifted to the
respective minimal values $1/2$ and
$1+\log \pi$, respectively, for $\beta=1$ (c), $\beta=2$ (d).
Illustration of the growing product of variances as functions of the classical parameter
$\mu$ when compared with the sum of entropies. }
\label{fig3}
\end{figure}


\begin{figure}[h]
\includegraphics[scale=0.7]{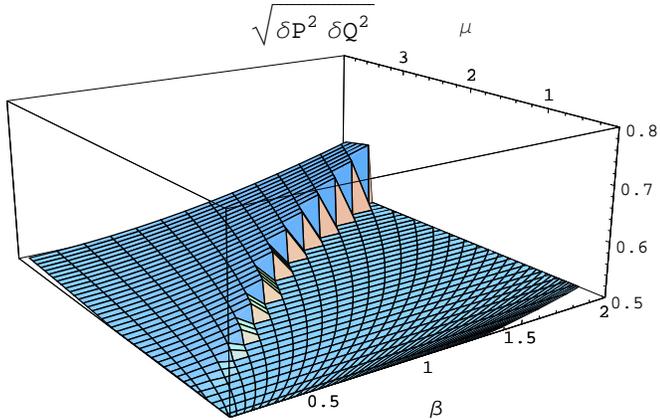}
\caption{Product of variances $\sqrt{\Delta P^2\Delta Q^2}$ calculated
from the variational approach of Section 3.1 (\ref{VA2}-\ref{VA3}). Compared 
to Fig.3 
the fluctuations are completely suppressed on the right part of the Figure.
 }
\label{fig4}
\end{figure}


\begin{figure}[h]
\includegraphics[scale=1]{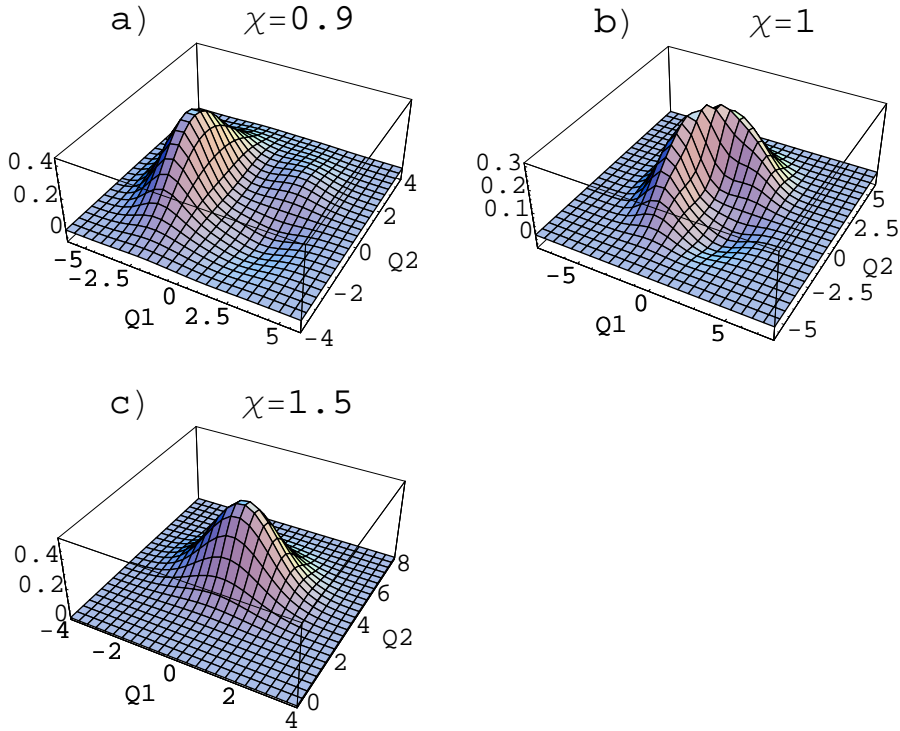}
\caption{The two-phonon numerical ground state wave functions at $\mu= 2$ and
$\chi=0.9$ (a), $\chi=1$ (b) and $\chi=1.5$ (c). }
\label{fig5}
\end{figure}



\begin{figure}[h]
\includegraphics[scale=0.85]{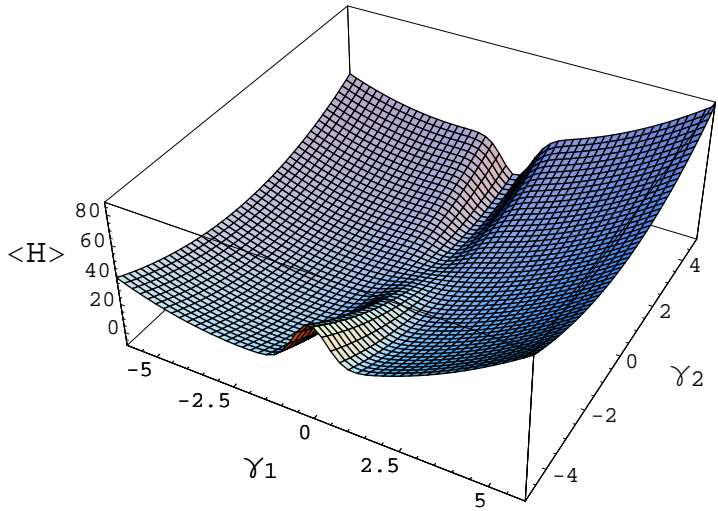}
\caption{Effective potential in the plane $\gamma_1$,
$\gamma_2$ containing several competing minima for $\mu=2$, $\chi=1.5$.
For this values of the parameters the narrow (tunneling) minimum
$\gamma_1\simeq 0,\gamma_2>0$ dominates. For $\chi<1$ the broad minimum at
$\gamma_1<0, \ \gamma_2\simeq 0$ would dominate (the selftrapping).  }
\label{fig6}
\end{figure}


\begin{figure}[h]
\includegraphics[scale=0.7]{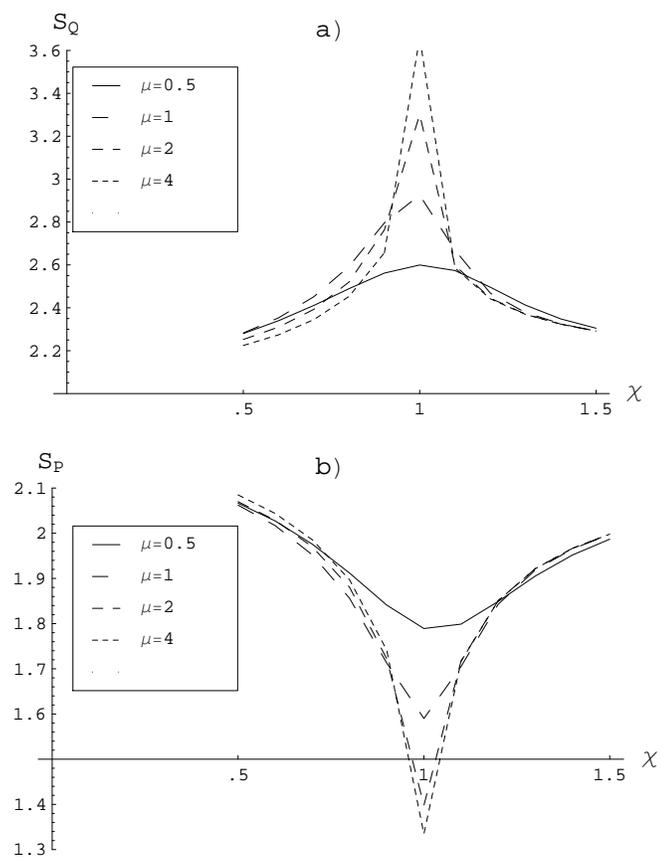}
\caption{Shannon entropies $S_{Q}$ (a) and $S_{P}$ (b) as functions
of $\chi$ and $\mu$. }
\label{fig7}
\end{figure}


\begin{figure}[h]
\includegraphics[scale=0.7]{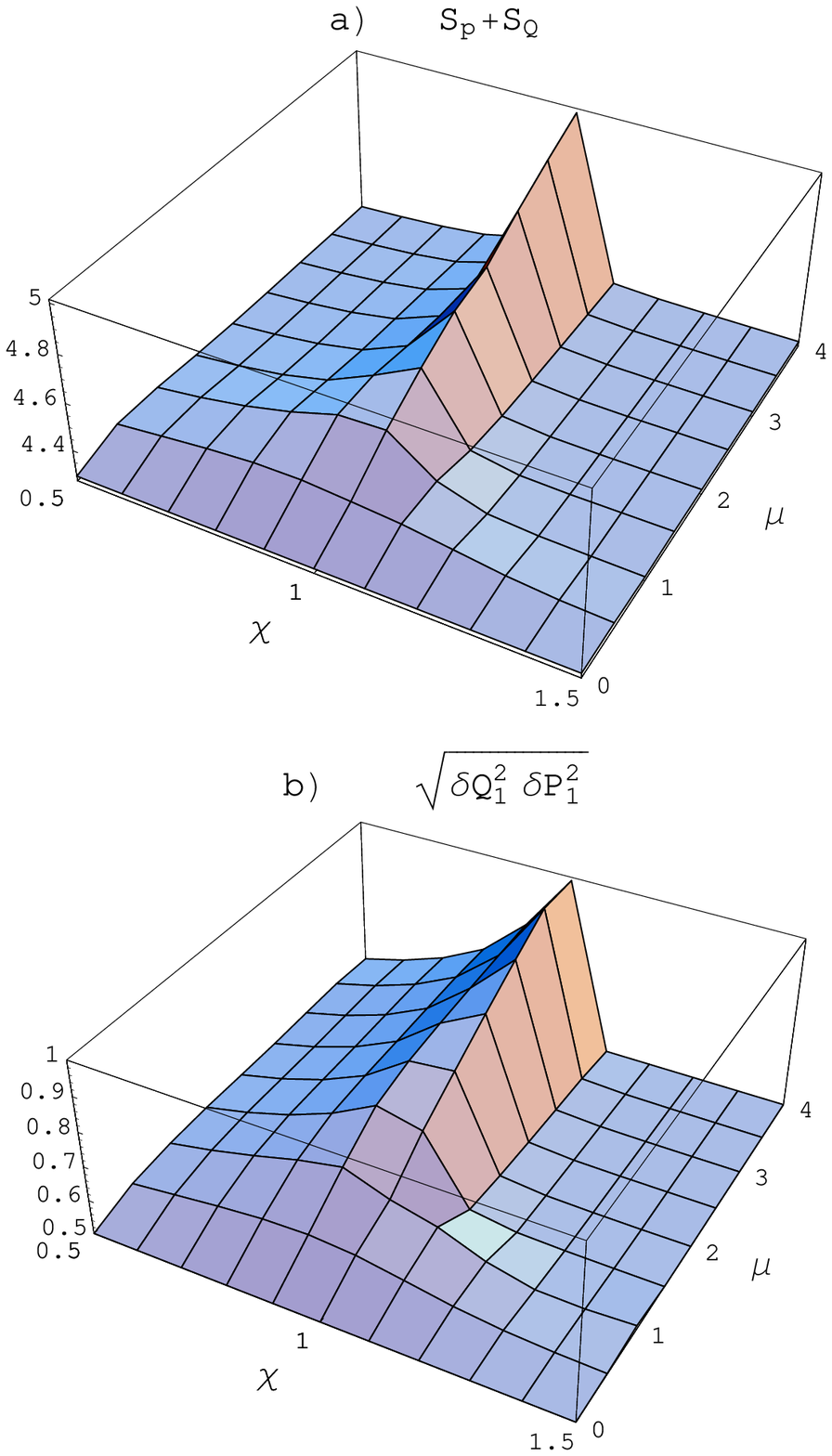}
\caption{Sum of entropies $S_{Q}+S_{P}$ in the plane $\chi,
\mu$. The lower bound corresponds to the two harmonic oscillators value
$2(1+\log \pi)$ (a); Product of variances $\sqrt {\Delta Q_1^2 \Delta P_1^2}$
 in the plane $\chi, \mu$ (b).} The difference of two Figs is not so well pronounced 
as on its one-dimensional counterpart Fig.3, but for better insight the reader 
is referred to following Fig.9 with pertaining crossections.
\label{fig8}
\end{figure}


\begin{figure}[h]
\includegraphics[scale=1.05]{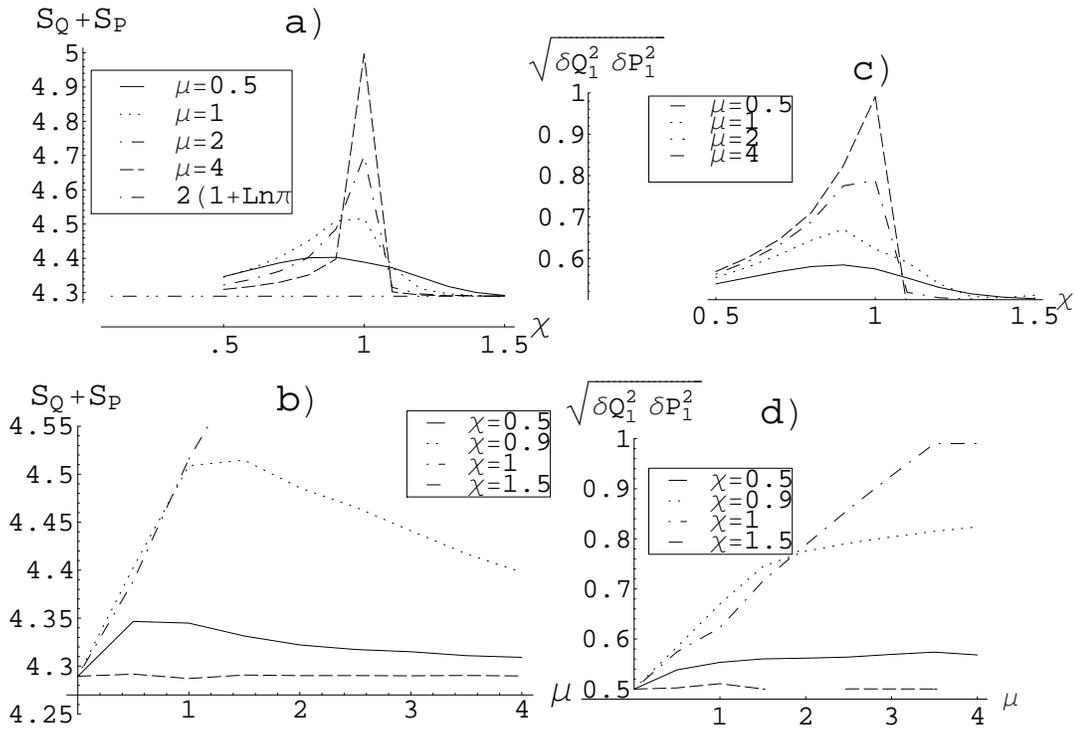}
\caption{Cross-section of the sum of entropies at Fig. 8a,
$S_{Q}+S_{P}$ as a function of $\chi$ for various $\mu$ (a).
The same as a function of $\mu$ for several $\chi$ (b).
Cross-section of the product of variances
$\sqrt{\Delta Q_1^2\Delta P_1^2 }$ at Fig. 8b as a function of $\chi$ for
various $\mu$ (c).
The same as a function of $\mu$ for various $\chi$ (d).}
\label{fig9}
\end{figure}


\begin{figure}[h]
\includegraphics[scale=0.7]{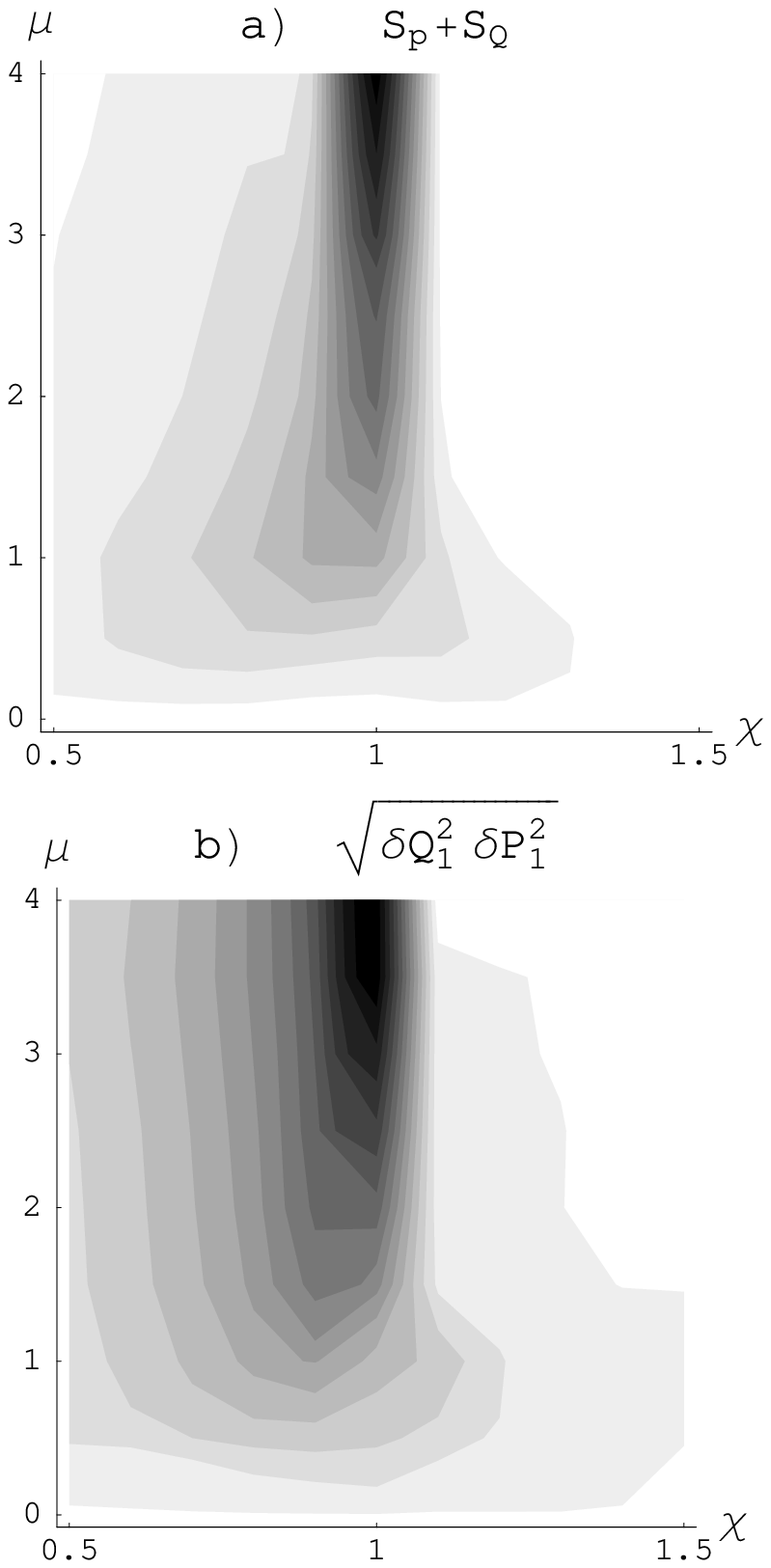}
\caption{ Contour plots of the sum of entropies of Fig. 8a
(a) and of the product of variances of Fig. 8b  (b). }
\label{fig10}
\end{figure}


\begin{thebibliography}{10}
\bibitem{descr} It is worth noting that 
the problem mentioned arises in more general context, namely within the realm of 
descriptive statistics. The conventional parametric statistics which is based on 
momenta estimation often fails to represent adequately the characteristics of 
nonuniform data and sometimes it leaves the scene in favour to the 
``nonparametric statistics''.
As the simplest example, consider the ordinary mean (first momentum) $\int x p(x) d x$ 
of a strongly 
asymmetric distribution function. Within the nonparametric statistics in order 
to characterize intuitive notion of the ``distribution center'' it is common to use 
the median, or $1/2$-quantile instead of the mean value.
\end{thebibliography}
\end{document}